\documentclass{ws-procs9x6}

\begin{document}

\title{Anomaly Mediated SUSY Breaking from a String Theory Perspective}

\author{Brent~D.~Nelson}

\address{Michigan Center for Theoretical Physics \\
University of Michigan \\ 3444A Randall Laboratory\\ Ann Arbor,
Michigan 48109}

\maketitle

\abstracts{The general form of soft supersymmetry breaking terms
arising from the superconformal anomaly are presented and are
examined in the context of effective field theories derived from
the weakly-coupled heterotic string. Some phenomenological
consequences are considered and compared to scenarios with
universal soft terms.}

There has been much interest of late in contributions to soft
supersymmetry-breaking parameters arising at one loop order from
the superconformal anomaly of supergravity.\cite{GiLuMuRa98} If
the superconformal anomaly is the {\em sole} contributor to soft
supersymmetry (SUSY) breaking then the resulting soft Lagrangian
shows a remarkable degree of insensitivity to high-energy physics,
perhaps providing a resolution to the so-called SUSY flavor
problem. Such a situation has been speculated to arise in certain
brane-world constructions that have been denoted ``sequestered
sector'' models.\cite{RaSu99} The soft supersymmetry breaking
parameters in this scenario are typically computed using a flat
space spurion technique\cite{PoRa99} which leads to the conclusion
that scalar masses are vanishing at the one loop level and arise
only at two loops. Since this leads to negative squared masses for
scalar leptons, some additional model building must be undertaken
to make this scenario phenomenologically viable.

The above description is not, however, the only possible
manifestation of ``anomaly-mediation'' of SUSY breaking but
represents a special case of a more general phenomenon. The
superconformal anomaly is an integral part of supergravity and its
contributions to soft SUSY breaking are truly ``gravity
mediation'' {\em par excellence}, taking their place among all
other contributions at the loop level to the soft Lagrangian. Thus
they will be present in the standard effective Lagrangians derived
from the weakly-coupled heterotic string, and not solely in those
based on strings at strong coupling.

What's more, a deeper understanding of the relationship between
chiral spurions in flat space and a true supergravity loop
calculation reveals that the former technique is
incomplete.\cite{GaNe00b} Indeed, the vanishing of scalar masses
at one loop is seen to be an artifact of certain (often tacitly
made) assumptions about the form of the K\"ahler potential and the
way in which the supergravity theory is ultimately regulated. That
the regularization of the theory enters the low-energy physics
indicates that the degree to which SUSY breaking is insensitive to
UV physics is a model dependent question.

To be more specific we can investigate the form of the soft SUSY
breaking terms in a modular invariant low-energy effective theory
defined by the K\"ahler potential $K(S,T) =
-\ln\left(S+\overline{S}\right)-3\ln\left(T+\overline{T}\right)$
with $S$ the chiral dilaton superfield and $T$ representing an
overall K\"ahler modulus. The K\"ahler metric for matter
superfields $Z^i$ will be taken to be $K_{i \bar{\jmath}} =
\kappa_i(T) \delta_{i\bar{\jmath}} + O(|Z^i|^2)$, with
$\kappa_i(T) = \left(T+\bar{T}\right)^{n_i}$, where $n_i$ is the
modular weight of the field $Z^i$. Modular invariance of the
Lagrangian is enforced by associating a moduli dependence with the
Yukawa couplings of the superpotential $W_{ijk} = w_{ijk} \left[
\eta(T) \right]^{-2(3+n_i+n_j+n_k)}$ where $\eta(T)$ is the
Dedekind function.

The contribution to gaugino masses $M_a$ arising at one loop from
the superconformal anomaly are given by\cite{GaNeWu99,BaMoPo00}
\begin{equation} M^{1}_a|_{\rm an} =
\frac{g_{a}^{2}(\mu)}{2} \left[ \frac{2 b_a}{3} \overline{M} -
2b_a' F^n  K_{n} - \frac{1}{4\pi^2} \sum_i C^i_a F^n \partial_n
\ln \kappa_i \right], \label{Man}
\end{equation}
where $C_a$, $C_a^i$ are the quadratic Casimir operators for the
gauge group $G_a$ in the adjoint representation and in the
representation of $Z^i$, respectively, and we have defined the
quantities $b_{a}$ and $b_{a}'$ by:
\begin{equation}
b_{a} = \frac{1}{16\pi^2} \left(3C_a - \sum_i C_a^i \right) ;
\quad b_{a}' = \frac{1}{16\pi^2} \left(C_a - \sum_i C_a^i \right)
. \label{ba}
\end{equation}
When combined with loop contributions from the Green-Schwarz
counterterm and string threshold corrections to the gauge kinetic
function, the complete tree plus one loop gaugino mass is
\begin{equation}
M_{a} = \frac{g_{a}^{2}(\mu)}{2} \left\{ 2
 \left[ \frac{\delta_{\rm GS}}{16\pi^{2}} + b_{a}
\right]G_{2}(t,\bar{t}) F^{T} + \frac{2}{3}b_{a}\overline{M}
+\left[ 1 - 2 b_{a}' K_s \right] F^{S} \right\} , \label{Maloop}
\end{equation}
where $F^S$ and $F^T$ are the auxiliary fields for the dilaton and
the K\"ahler modulus, respectively, $M$ is the supergravity
auxiliary field whose vacuum expectation value (vev) determines
the gravitino mass $m_{3/2} = -\frac{1}{3}\langle \overline{M}
\rangle= \langle e^{K/2} \overline{W} \rangle$, and $\delta_{\rm
GS}$ is the Green-Schwarz coefficient which is a (negative)
integer between 0 and -90. The function $G_{2}(t,\bar{t})$ is
proportional to the Eisenstein function and vanishes when $T$ is
stabilized at one of its two self-dual points. From~(\ref{Maloop})
it is clear that when supersymmetry breaking is communicated to
the observable sector via a non-zero vev for $F^T$ while the
moduli are stabilized at a self-dual point then only the second
term contributes to gaugino masses. This is precisely the
``anomaly mediated'' contribution.\cite{GiLuMuRa98,RaSu99}

The soft supersymmetry breaking parameters that appear in the
scalar potential have a more complicated form at one loop in
supergravity and will depend on the way in which the theory is
regulated. These loop contributions have been computed
recently\cite{GaNe00b} using the Pauli-Villars (PV) regularization
procedure.\cite{Ga95} To understand their basic structure it is
sufficient to make the simplifying assumption that all of the
observable sector matter fields have modular weights $n_i = -1$.
This assumption also dictates the modular weights of the
Pauli-Villars fields that regulate the loop contributions from the
light fields, but not the modular weights of the fields that
generate the (supersymmetric) Pauli-Villars masses themselves. We
therefore introduce a phenomenological parameter $p$ that
represents the effective modular weights of these fields.

With these simplifications the relevant soft terms are given by
\begin{eqnarray}
A_{ijk} &=& -\frac{K_s}{3}F^S - \frac{1}{3} \gamma_{i}\overline{M}
- p \gamma_{i} G_{2}(t,\bar{t}) F^{T} + \tilde{\gamma}_{i} F^{S}
\ln\left(\widetilde{\mu}_{\rm PV}^{2}/\mu_R^2\right) + {\rm
cyclic}(ijk) \nonumber \\
M_{i}^{2} &=& \left\{\frac{|M|^2}{9}
-\frac{|F^T|^{2}}{(t+\bar{t})^{2}}\right\} \left[ 1 + p\gamma_i
-\left(\sum_{a}\gamma_{i}^{a} -2\sum_{jk}\gamma_{i}^{jk}\right)
\ln\left(\widetilde{\mu}_{\rm PV}^{2}/\mu_R^2\right) \right]
\nonumber \\
& & + (1-p)\gamma_i \frac{|M|^2}{9} +
\widetilde{\gamma}_{i}\frac{F^S}{2}\left\{\frac{M}{3}+ p\;
G_{2}(t,\bar{t}) \overline{F}^{T} + \mathrm{\; h.c. \;} \right\}
\nonumber \\
& & +|F^{S}|^2 \left[ \left(\frac{3}{4}\sum_{a} \gamma_{i}^{a}
g_{a}^{2} + K_s K_{\bar{s}} \sum_{jk} \gamma_{i}^{jk}\right)
 \ln\left(\widetilde{\mu}_{\rm PV}^{2}/\mu_R^2\right) \right],
\label{finalsoft}
\end{eqnarray}
where the quantity $ \ln\left(\widetilde{\mu}_{\rm
PV}^{2}/\mu_R^2\right)= \ln\left(\mu_{\rm PV}^{2}/\mu_R^2\right)
-p\ln\left[(t+\bar{t}) |\eta(t)|^4\right]$ can be interpreted as a
threshold correction to the Pauli-Villars mass scale $\mu_{\rm
PV}$ and $\mu_R$ being the renormalization scale. The various
quantities related to the anomalous dimensions $\gamma_i^j \approx
\gamma_i \delta_i^j = \sum_{jk}\gamma_i^{jk} + \sum_a\gamma_i^a$
of the fields are defined by
\begin{equation}
\gamma_i^a = \frac{g^2_a}{8\pi^2}(T^2_a)^i_i, \quad \gamma_i^{jk}
= -\frac{e^K}{32\pi^2}
(\kappa_i\kappa_j\kappa_k)^{-1}\left|W_{ijk}\right|^2.\label{gamma}
\end{equation}
and $\widetilde{\gamma}_i$ is a shorthand notation for the
quantity $\widetilde{\gamma}_{i} = \sum_{a} \gamma_{i}^{a}
g_{a}^{2} - K_{s} \sum_{jk} \gamma_{i}^{jk}$.

The expressions in~(\ref{Maloop}) and~(\ref{finalsoft}) represent
the complete one loop expressions. As mentioned above, the case
where the anomaly-mediated terms proportional to $M$ are dominant
is the case of moduli domination $\langle F^S \rangle =0$,
$\langle F^T \rangle \neq 0$ with the moduli stabilized at a
self-dual point. In that case, if the vacuum energy vanishes the
tree-level scalar masses are zero and the soft terms are given by
\begin{equation}
M_{a}=g_{a}^{2}(\mu) b_{a}\frac{\overline{M}}{3}\; ; \; A_{ijk}=
-(\gamma_{i}+\gamma_{j}+\gamma_{k}) \frac{\overline{M}}{3}\; ; \;
M_{i}^{2}=(1-p)\gamma_{i}\frac{|M|^{2}}{9}.
\end{equation}
The parameter $p$ is constrained to be no larger than 1, though it
can be negative in value. Thus {\em the scalar squared mass for
all matter fields will in general be non-zero and positive at one
loop}~-- only the Higgs can have a negative running squared mass.
The limiting case of $p=1$, where the scalar masses are zero at
the one loop level and we recover the sequestered sector limit,
occurs when the regulating PV fields and the mass-generating PV
fields have the same dependence on the K\"ahler moduli. Another
reasonable possibility is that the PV masses are independent of
the moduli, in which case we would have $p=0$.

Even when the anomaly-mediated terms are not the sole contributors
to the soft supersymmetry-breaking Lagrangian they can still have
an important influence over the resulting
phenomenology.\cite{BiGaNe01} This influence is most pronounced in
the gaugino mass sector, where one often encounters vanishing or
suppressed tree level contributions to the gaugino masses in
string-inspired models of supersymmetry breaking.\cite{benchmarks}
The resulting interplay between the various gauge group dependent
contributions in~(\ref{Maloop}) can then give rise to gluinos that
are lighter than would be expected in a unified gaugino mass
scenario, while still providing a heavy enough chargino to avoid
the LEP direct search limit. This can have significant
implications for fine-tuning in the electroweak
sector\cite{finetune} and the thermal relic abundance of the
lightest neutralino.\cite{BiNe01,BiNe02}


\end{document}